\documentclass{iau_FM}
\usepackage{graphicx}

\title[JD 8.~~Magnetic massive stars] %% give here short title %%
{Magnetic massive stars:\\confirming the merger scenario\\for the magnetic field generation}

\author[S.~Hubrig et al.]   %% give here short author list %%
{Swetlana Hubrig$^1$, Markus~Sch{\"o}ller$^2$, Silva~P.~J\"arvinen$^1$, \\
Aleksandar Cikota$^3$, Michael Abdul-Masih$^4$,
Ana Escorza$^4$, Ilya Ilyin$^1$}
\affiliation{$^1$Leibniz-Institut f\"ur Astrophysik Potsdam (AIP), An der Sternwarte~16, 14482~Potsdam, Germany\\
email: {\tt shubrig@aip.de, sjarvinen@aip.de, ilyin@aip.de} \\[\affilskip]
$^2$European Southern Observatory, Karl-Schwarzschild-Str.~2, 85748~Garching, Germany\\
email: {\tt mschoell@eso.org} \\[\affilskip]
$^3$Gemini Observatory / NSF's NOIRLab, Casilla 603, La~Serena, Chile\\ email: {\tt aleksandar.cikota@gmail.com} \\[\affilskip]
$^4$Instituto de Astrof\'isica de Canarias, C. V\'ia L\'actea s/n, 38205~La~Laguna, Santa~Cruz~de~Tenerife, Spain\\
email: {\tt michael.abdul-masih@iac.es, ana.escorza.santos@iac.es }}

\pubyear{2024}
\jname{Astronomy in Focus, Focus Meeting 8} 
\editors{Alexander Kosovichev, ed.}
\begin{document}

\maketitle

\begin{abstract}
Magnetic fields are considered to be key components of massive stars, with a
far-reaching impact on their evolution and ultimate fate.
A magnetic mechanism was suggested for the collimated explosion of massive stars, relevant for
long-duration gamma-ray bursts, X-ray flashes, and asymmetric core collapse supernovae.
However, the origin of the observed stable, globally
organized magnetic fields in massive stars is still a matter of debate: it has been argued that they can
be fossil, dynamo generated, or generated by strong binary interactions or merging events.
Taking into account that  multiplicity is a fundamental characteristic of
massive stars, observational evidence is accumulating that the magnetism originates through 
interaction between the system components, both during the initial mass transfer
or when the stellar cores merge.
\end{abstract}

\firstsection % if your document starts with a section,
\section{Introduction: Scientific background}

\begin{figure*}
  \begin{center}
   \includegraphics[width=2.33in]{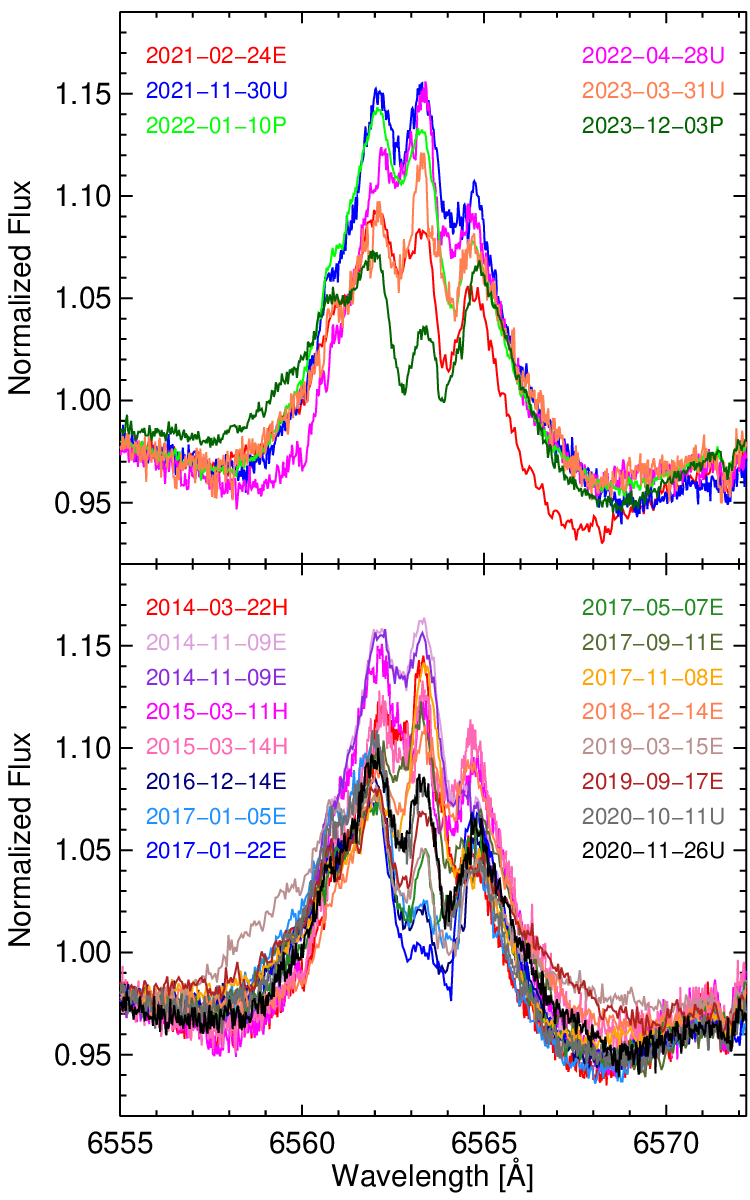} 
 \includegraphics[width=2.87in]{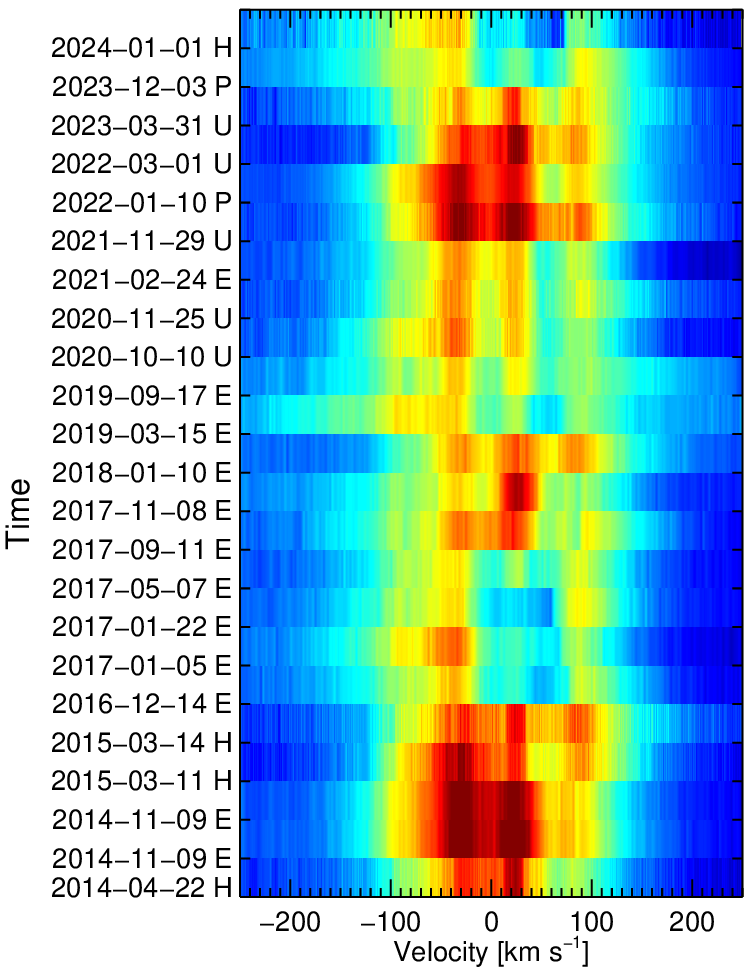}
 \caption{
   {\it Left:} Overplotted H$\alpha$ line profiles observed in the magnetic O-type star HD\,54879
  with an extremely long rotation period of 7.4\,yr.
Each profile is plotted with a different color and the instrument used is
indicated after each date: HARPS (H), ESPaDOnS (E), UVES (U), and PEPSI (P).
The upper panel shows the profiles based on the most recent spectra and the bottom
panel shows the profiles already published in 
\cite[J{\"a}rvinen et al.\ (2022)]{Silva2022}.
{\it Right:} Dynamical spectrum based on the observed
variability of the H$\alpha$ profiles.
The red colour corresponds to the strongest H$\alpha$ emission appearing at the time of the best visibility of
the dipole magnetic poles.
}
   \label{fig1}
\end{center}
\end{figure*}

Unlike Sun-like stars, which possess magnetic fields with complex topologies generated through a dynamo process 
sustained by a convective dynamo operating close to the
surface of the star, magnetic massive stars host large-scale organized
fields with simple topologies usually described by a magnetic dipole tilted to the rotation axis.
Stellar winds in massive magnetic stars are confined by the magnetic field building dynamical or
centrifugal magnetospheres: slowly rotating stars for which the Alfv\'en radius $R_A$ is smaller than the
Kepler radius $R_K$  build dynamical magnetospheres and more rapidly rotating stars with $R_A>R_K$
build centrifugal magnetospheres 
(\cite[Petit et al.\ 2013]{Petit2013}).
In Fig.~\ref{fig1}
we present the first image of the dynamical
magnetosphere around a typical magnetic O star, the slowly rotating ($P_{\rm rot}=7.4$\,yr) O star HD\,54879 with a dipolar magnetic field
of kG-order. This image is produced using monitoring of the H$\alpha$ profile over 9.7\,yr 
(\cite[K{\"u}ker et al.\ 2024]{Kueker2024}).

Studies of the magnetic characteristics of massive stars have recently received significant attention
because these stars are progenitors of highly magnetised compact objects. Stars initially
more massive than about $8\,M_{\odot}$ leave behind neutron stars and black holes by the end of their
evolution. The merging of binary compact remnant systems produces astrophysical transients
detectable by the gravitational wave observatories LIGO, Virgo, and KAGRA.
Also studies of magnetic fields in massive stars in a low metallicity environment are of particular interest because
they provide important information on the role of magnetic fields in the star formation of the early Universe.
The recent discovery of magnetic fields of kG-order in three massive O-type targets in the Magellanic Clouds, two stars with
spectral classification Of?p and one overcontact binary, suggests that the impact of low metallicity on
the occurrence and strength of magnetic fields in massive stars cannot be strong 
(\cite[Hubrig et al.\ 2024]{Hubrig2024}).

Despite the progress achieved in previous surveys of the magnetism in massive stars,
the origin of their magnetic fields remains to be the least understood topic.
It has been argued that magnetic fields could be fossil, dynamo generated, or
generated by strong binary interaction.
The currently most popular theoretical scenarios developed to explain the origin of magnetic fields in massive stars
involve a merger event or a common envelope evolution 
(e.g.\
\cite[Tout et al.\ 2008]{Tout2008};
\cite[Ferrario et al.\ 2009]{Ferrario2009};
\cite[Schneider et al.\ 2016]{Schneider2016}).
Mass transfer or
stellar merging may rejuvenate the mass gaining star, while the induced differential rotation is
thought to be the key ingredient to generate a magnetic field 
(e.g.\
\cite[Wickramasinghe et al. 2014]{Wickramasinghe2014}).
Such interaction can take place already 
in the early evolutionary stages in star forming regions where accretion-driven  migration as well as tides induced by
the dense circumbinary material can lead to shrinking orbits and merging.

In the context of this scenario, O stars are of special interest to support these theoretical considerations as they
form nearly exclusively in multiple systems, with more than 90\% born and living in such
systems 
(\cite[Offner et al.\ 2023]{Offner2023}).
\cite[Moe \& di Stefano (2017)]{MoediStefano2017}
reported that
the O-type multiplicity fraction is 6$^{+6}_{-3}$\% for single stars, 
$21\pm$7\% for binary systems, $35\pm$3\% for triple systems, and $38\pm$11\% for quadruple
systems, meaning that about 73\% of O stars are members of multiple systems. According to these authors, the mechanisms required to
bring the inner binary systems in multiple systems to shorter periods include migration through a circumbinary disk due
to hydrodynamical forces, dynamical interactions in an initially unstable hierarchical multiple
system, or secular evolution caused by the so-called Kozai-Lidov mechanism
(\cite[Kozai 1962]{Kozai1962};
\cite[Lidov 1962]{Lidov1962}).
\cite[Sana et al.\ (2012)]{Sana2012}
reported that in the course of the stellar evolution $71\pm$8\% of O-type stars will interact
with companions in the systems
with mass ratios $q>$0.1 via Roche lobe overflow. This fraction should be even larger if the variations between orbital
periods, q, and eccentricities are taken into account 
(\cite[Moe \& di Stefano 2017]{MoediStefano2017}).

\section{Recent developments}

Curiously, in spite of the fact that the evolution of massive stars is highly affected by interactions in binary and multiple systems,
previous surveys of the magnetism in massive
stars indicated that the incidence rate of magnetic components in close binaries with $P_{\rm orb}< 20$\,d is very low:
of the dozen O stars with confirmed magnetic fields, only one short-period binary system, Plaskett's star, contains
a hot, massive, magnetic star 
(\cite[Grunhut et al.\ 2013]{Grunhut2012}).
However, it must be admitted that several factors hinder 
magnetic field detections in multiple systems and should be taken into account: the amplitudes of the
Zeeman features (features appearing in the Stokes~$V$ spectra of magnetic stars)
are much lower in multiple systems in comparison to their size in single stars. These features
also appear blended in composite spectra and show severe shape distortions.
Also the shapes of blended spectral lines in the Stokes~$I$ spectra look different depending on the visibility of each
system component at different orbital phases.
Given the much lower number of spectral lines in O stars in comparison to less massive stars, to measure their magnetic fields, the
least-squares deconvolution (LSD) technique 
(\cite[Donati et al.\ 1997]{Donati1997})
is frequently applied. However, special care has to be
taken to populate the LSD line masks for each system because the composite spectra of multiple systems usually show very different spectral
signatures corresponding to the different spectral classification of the individual components.

\begin{figure*}
\begin{center}
\includegraphics[width=1.6in]{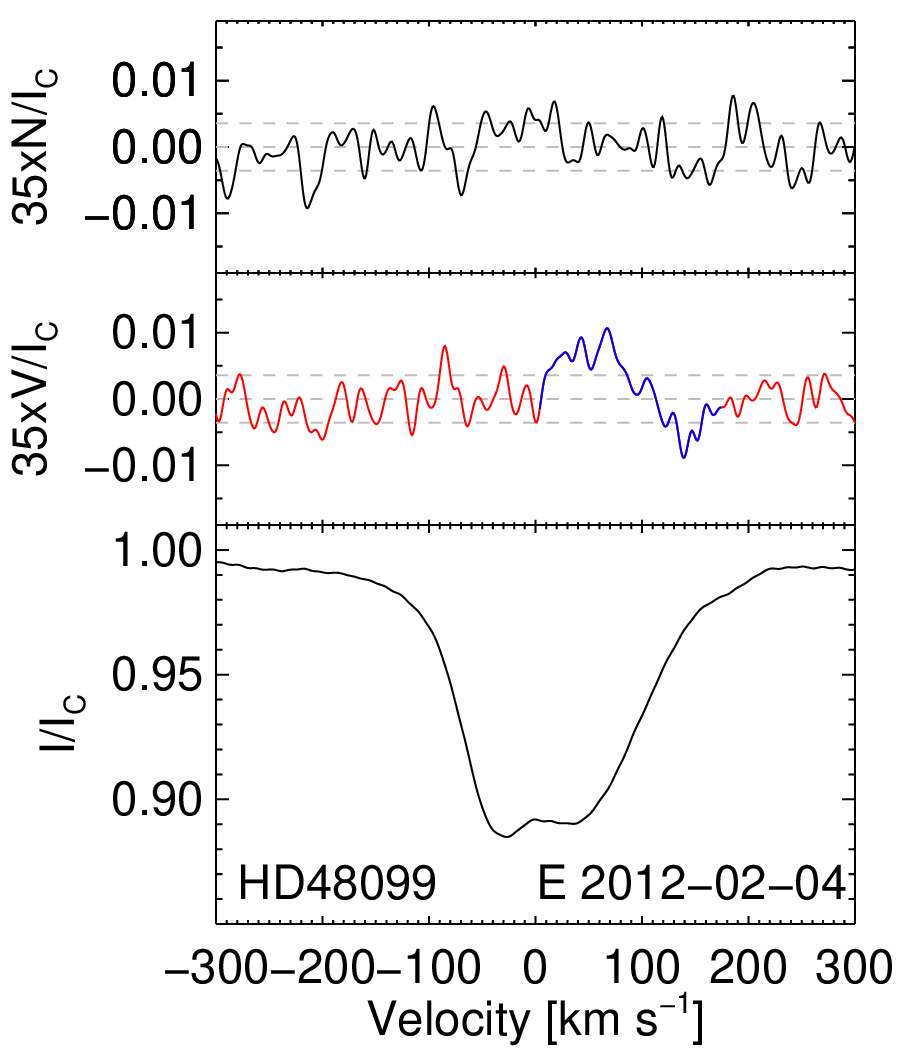}
\includegraphics[width=1.6in]{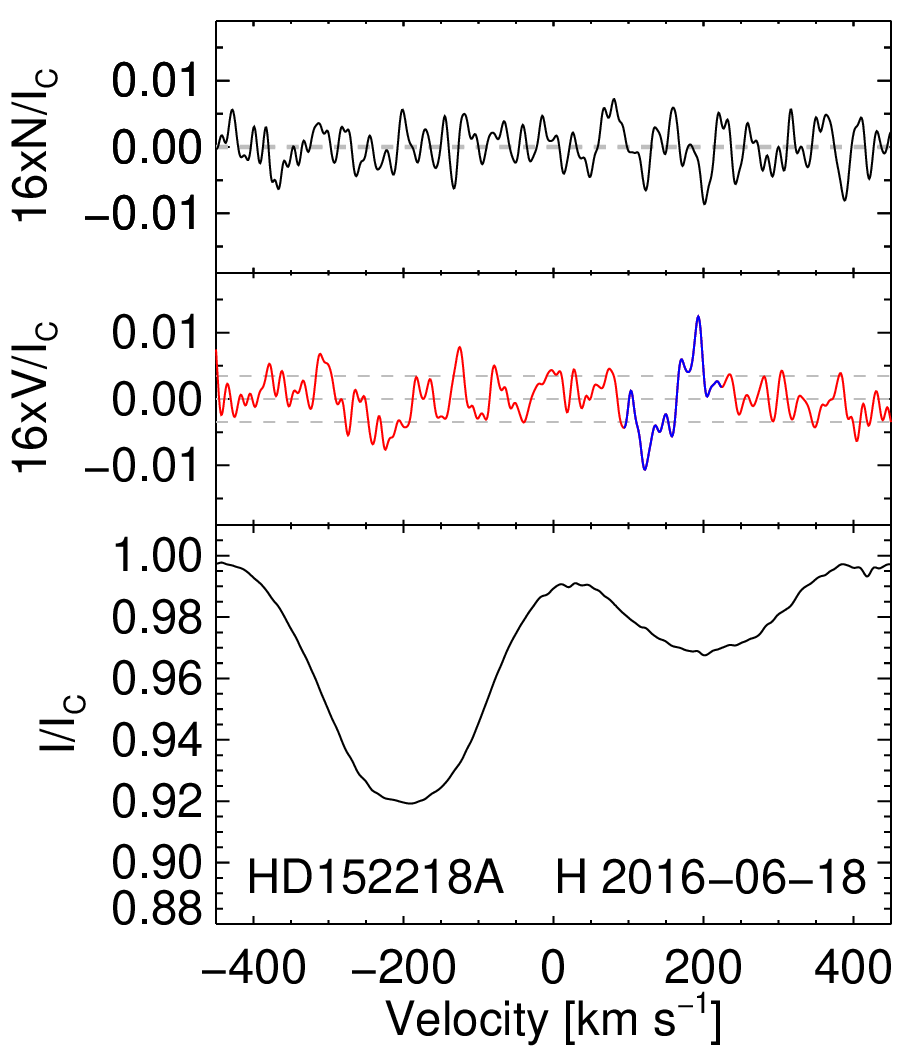}
\includegraphics[width=1.6in]{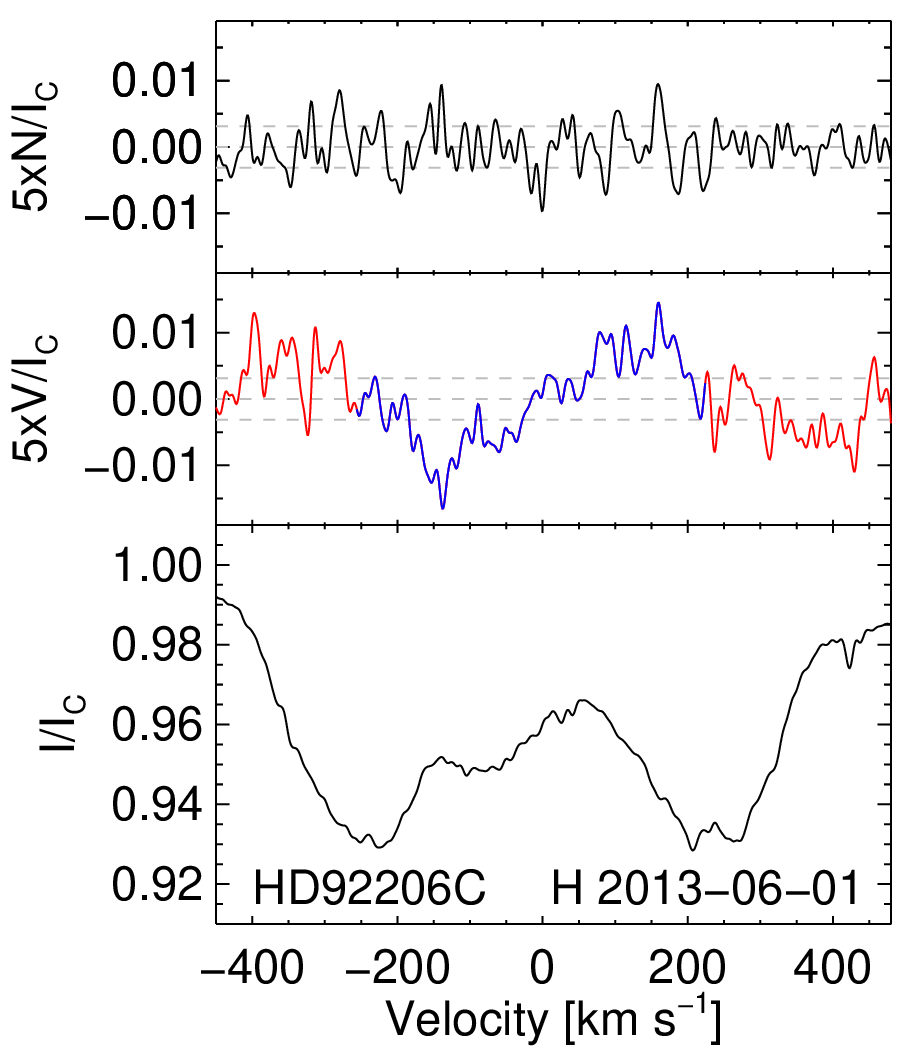}
\caption{
Examples of least square deconvolution Stokes~$I$, $V$, and diagnostic null $N$ spectra (from bottom to top)
for binary and multiple systems with recently discovered magnetic fields.
HD\,48099 is an X-ray colliding wind binary with a 3.1\,d orbital period 
(\cite[Mahy et al.\ 2010]{Mahy2010}).
HD\,152218A is an eclipsing eccentric binary with an orbital period of 5.6\,d
(\cite[Rosu et al.\ 2022]{Rosu2022}).
HD\,92206C is a triple system consisting of a SB2 system with an orbital period of 2.02\,d and a third component
at a distance of less than 1$^{\prime\prime}$ 
(\cite[Mayer et al.\ 2017]{Mayer2017}).
}
   \label{fig2}
\end{center}
\end{figure*}

Importantly, the first analysis of spectropolarimetric ESO and CFHT archival observations of 36 binary and multiple systems
with O-type primaries using a special procedure involving different line masks populated for each element separately
showed encouraging results 
(\cite[Hubrig et al.\ 2023]{Hubrig2023}):
out of the 36 systems, 22 exhibited in their least-squares deconvolution Stokes~$V$ profiles definitely detected Zeeman
features. For 14 systems, the detected Zeeman features have been associated with
O-type components, whereas for 3 systems they were associated with an early B-type component.
This survey included systems at very different evolutionary
stages, from young main-sequence systems to a few evolved systems with blue supergiants, Wolf-Rayet stars,
one system with a Luminous Blue Variable candidate, and one post-supernova X-ray binary.
Seven systems with definite field detections are known as
particle-accelerating colliding-wind binaries exhibiting synchrotron radio emission. A few examples of binary and multiple systems
with discovered magnetic fields are presented in Fig.~\ref{fig2}.

Strong indications that interaction between the components in binary and multiple systems can be important for the generation of
magnetic fields comes also from a recent study of the massive Of?p binary HD\,148937 
(\cite[Frost et al.\ 2024]{Frost2024}),
in which the
primary is suggested to be a merger product. Numerical simulations also indicate the important role of stripped-envelope stars formed in
systems with interacting components through one or multiple phases of Roche-lobe overflow 
(e.g.\
\cite[Yoon et al.\ 2017]{Yoon2017}).
Magnetic studies of such systems are of special interest because
these systems occupy the mass range that has been predicted to produce most stripped-envelope supernovae or
neutron star mergers like those that emit gravitational waves 
(\cite[Tauris et al.\ 2017]{Tauris2017};
\cite[G{\"o}tberg et al.\ 2023]{Gotberg2023}).
Recent detections of magnetic fields in stripped stars have been reported by 
\cite[Hubrig et al.\ (2022)]{Hubrig2022}
for $\upsilon$~Sgr  and by 
\cite[Shenar et al.\ (2023)]{Shenar2023}
for HD\,45166.

\begin{figure*}
\begin{center}
\includegraphics[width=1.7in]{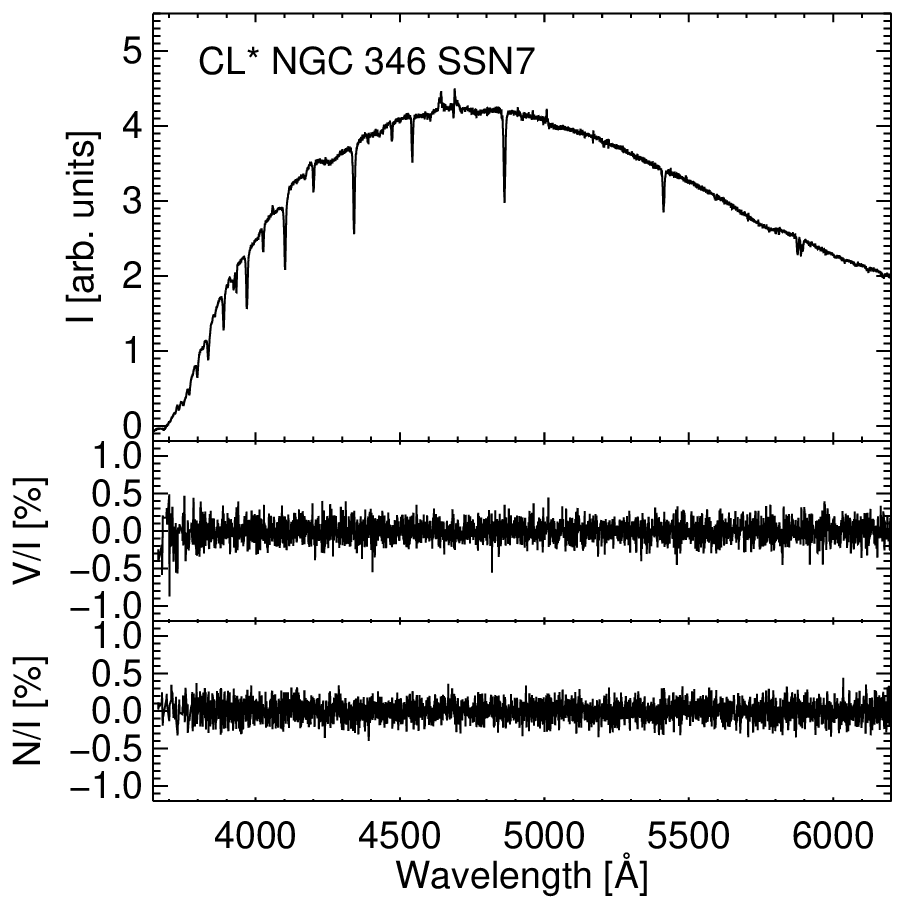}
\includegraphics[width=1.8in]{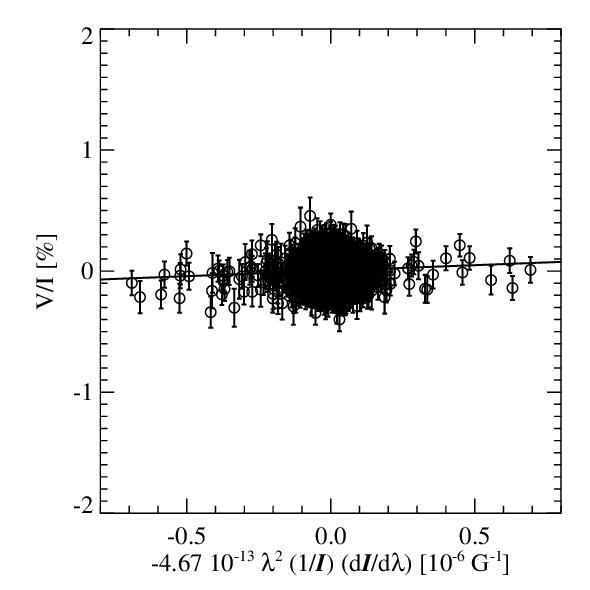}
 \caption{
FORS\,2 observations of the overcontact binary Cl*\,NGC\,346\,SSN\,7.
In the left panel, from top to bottom, the solid lines show the recorded Stokes~$I$ and
Stokes~$V$ spectra and the diagnostic $N_V$ spectra.
In the right panel, we show the regression determination of the mean longitudinal magnetic field
 $\left<B_{\rm z}\right>_{\rm all}=908\pm206$\,G
detected at a 4.4\,$\sigma$ significance level.
   }
   \label{fig3}
\end{center}
\end{figure*}

Perhaps the most interesting piece of
observational evidence for the important role of component interaction for the generation of magnetic fields
comes from the detection of rather strong longitudinal magnetic fields in three
massive overcontact binaries. In our Galaxy, magnetic fields have been detected in HD\,35921 (=LY\,Aur) and  HD\,167971 (=MY\,Ser)
(\cite[Hubrig et al.\ 2023]{Hubrig2023}).
Most recently, a magnetic field of kG-order has been reported for  
Cl*\,NGC\,346\,SSN\,7 in the Small Magellanic Clouds 
(\cite[Hubrig et al.\ 2024]{Hubrig2024}).
This was the first reported magnetic field in
an extragalactic overcontact binary system. In Fig.~\ref{fig3} we present the recent discovery of the magnetic field in the system
Cl*\,NGC\,346\,SSN\,7 based on low-resolution spectropolarimetric observations using the FORS2 instrument attached to an ESO VLT 8m telescope.
The mean longitudinal magnetic field from the FORS\,2 observations is usually determined as the slope of a weighted
linear regression through the measured Stokes~$V$ values.
The detection of magnetic fields in a representative number of overcontact systems, which directly
precede the merger event, would give further credence to the theoretical scenario presented by 
\cite[Schneider et al.\ (2019)]{Schneider2019},
who carried out three-dimensional magnetohydrodynamical simulations of the coalescence of two massive stars.

\section{New directions}

\begin{figure*}
\begin{center}
\includegraphics[width=1.6in]{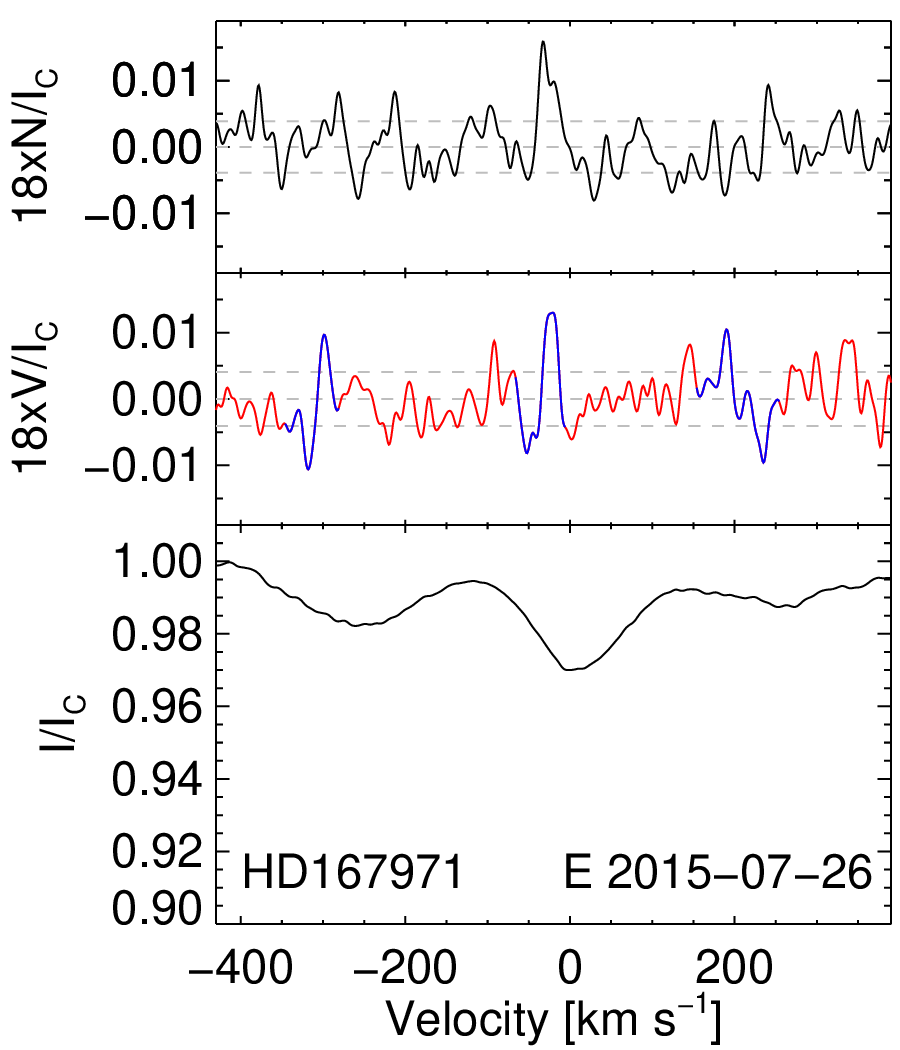}
\includegraphics[width=1.6in]{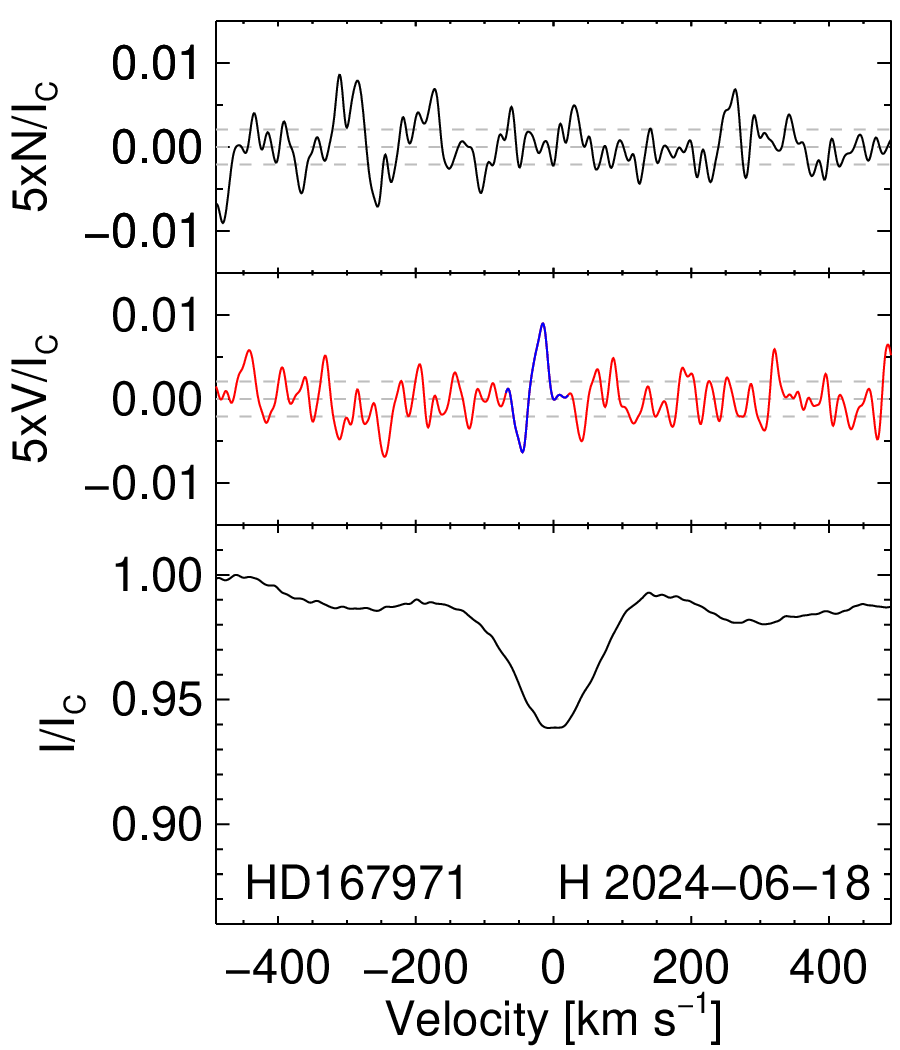}
\includegraphics[width=1.6in]{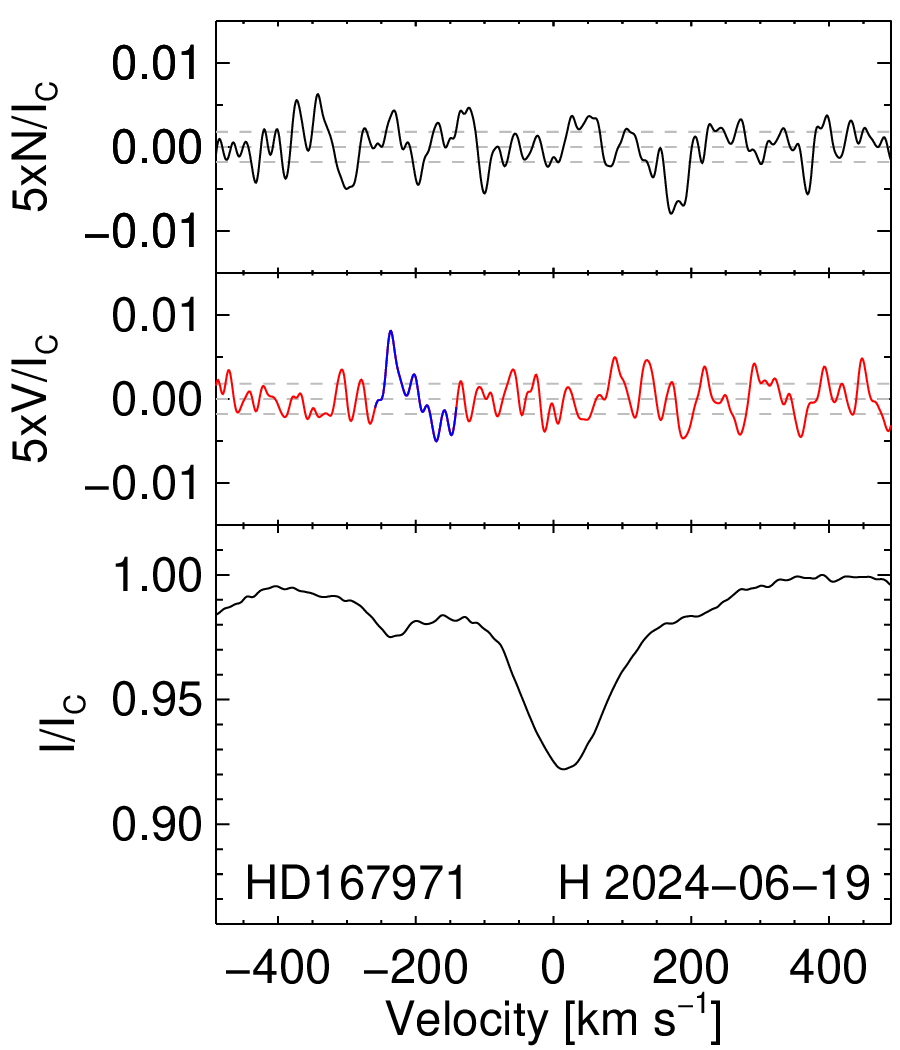}
 \caption{
   Examples of previous and most recent observations of the magnetic components in the triple system HD\,167971 (=MY\,Ser).
   HD\,167971 is one of the rare massive O-type triple systems in which the
secondary and the tertiary components compose
an overcontact eclipsing binary with an orbital period of 3.32\,d 
(\cite[Ibanoglu et al.\ 2013]{Ibanogl2013}).
   }
   \label{fig4}
\end{center}
\end{figure*}

The gradually increasing number of systems with definitely detected magnetic fields
indicates that multiplicity plays a crucial role in the generation of
magnetic fields in massive stars. A few systems with magnetic components studied by 
\cite[Hubrig et al.\ (2023)]{Hubrig2023}
have recently been re-observed with HARPS\-pol confirming their magnetic nature.
In Fig.~\ref{fig4} we show as an example previous and
new observations of the triple system HD\,167971. As most of the targets have been observed with
spectropolarimetry only once or twice, the newly found magnetic systems are supreme candidates for spectropolarimetric monitoring over their
orbital and rotation periods.

The number of spectropolarimetric observations of massive stars is still too low to allow us to confirm current theories and
simulations developed to explain the origin of magnetic fields in massive stars. 
In view of the mounting evidence for the importance of studying magnetic fields in massive binary and multiple
systems, it is crucial to obtain trustworthy statistics on the magnetic field incidence, the magnetic field structure, and the
field strength distribution in systems with magnetic components. Because not all magnetic components have known rotation periods,
spectropolarimetric time-series are needed to determine the field structure. 
We also need to know what fraction of O stars are magnetic and what is the field structure across the different evolutionary stages.
On the other hand, this requires the determination of fundamental properties and the evolutionary history of the magnetic components
in binary and multiple systems.
Only then can we try to understand the impact of the generated magnetic fields on the further
evolution and the final fate of massive stars as supernovae and compact objects.

\end{document}